\documentclass[a4paper]{article}

\usepackage{INTERSPEECH2022}
\usepackage{multirow}
 \usepackage{color} 
 \usepackage{bbding, subfigure, url, hyperref}
\title{Learning Noise-independent Speech Representation for High-quality Voice Conversion for Noisy Target Speakers}

\name{Liumeng Xue$^1$, Shan Yang$^2$, Na Hu$^2$, Dan Su$^2$, Lei Xie\thanks{Major work performed while Liumeng Xue interning at Tencent AI Lab. *Lei Xie is the corresponding author. This work was supported by the National Key R\&D Program of China (2020AAA0108600).}$^{1,*}$}

\address{
  $^1$Audio, Speech and Language Processing Group (ASLP@NPU), School of Computer Science, \\ Northwestern Polytechnical University, Xi'an, China\\
  $^2$ Tencent AI Lab, China}
\email{ \{lmxue, lxie\}@nwpu-aslp.org, \{shaanyang, ninahu, dansu\}@tencent.com}

\begin{document}

\maketitle
\vspace{-3pt}
\begin{abstract} \vspace{-3pt}
Building a voice conversion system for noisy target speakers, such as users providing noisy samples or Internet found data, is a challenging task since the use of contaminated speech in model training will apparently degrade the conversion performance. In this paper, we leverage the advances of our recently proposed Glow-WaveGAN~\cite{cong2021glow} and propose a noise-independent speech representation learning approach for high-quality voice conversion for noisy target speakers. Specifically, we learn a latent feature space where we ensure that the target distribution modeled by the conversion model is exactly from the modeled distribution of the waveform generator. With this premise, we further manage to make the latent feature to be noise-invariant. Specifically, we introduce a noise-controllable WaveGAN, which directly learns the noise-independent acoustic representation from waveform by the encoder and conducts noise control in the hidden space through a FiLM~\cite{perez2018film} module in the decoder. As for the conversion model, importantly, we use a flow-based model to learn the distribution of noise-independent but speaker-related latent features from phoneme posteriorgrams. Experimental results demonstrate that the proposed model achieves high speech quality and speaker similarity in the voice conversion for noisy target speakers.
\end{abstract}
\noindent\textbf{Index Terms}: voice conversion, noise, noise robustness

\vspace{-3pt}
\section{Introduction}  \vspace{-3pt}
Voice conversion (VC) aims at transforming the vocal timbre of the source speech to the target speaker while preserving its linguistic content. It has many applications, including movie dubbing~\cite{mukhneri2020voice}, speaking assistance~\cite{nakamura2012speaking} and singing~\cite{sisman2020generative, li2021ppg, wang2021towards}. With the advances of deep learning, neural voice conversion methods have been studied extensively in recent years with high-quality natural converted speech~\cite{mohammadi2017overview}, such as generative adversarial network (GAN)-based ~\cite{kameoka2018stargan, fang2018high}, variational autoencoder (VAE)-based~\cite{kameoka2018acvae}, autoencoder-based~\cite{qian2019autovc} and flow-based~\cite{serra2019blow} models, to name a few.

But in real applications, speech from both the source and the target speakers may inevitably contain environmental noise during speech recording, such as samples provided by users in noisy environment or from the Internet. Thus a voice conversion approach desires to be noise-robust. Existing noise-robust conversion approaches mostly aim at addressing the background noise existing in the source speech, with the premise to convert the noisy source speech to the target speakers with clean speech for system building~\cite{chan2021speech, aihara2014noise, takashima2013noise, miao2019noise}. In other words, the target speaker's speech samples are assumed to be recorded at a studio-quality level. It is also challenging to deal with the noisy target as the use of contaminated speech in model training will definitely affect the conversion quality. We notice that there are only a few studies dealing with noisy target speakers. For example, adversarial training is used to learn noise-invariant content and speaker representation for noise-robust voice conversation~\cite{du2022noise}. Besides, an intuitive way is to leverage a speech enhancement module to remove noise before training~\cite{huang2021toward}, but it will inevitably affect the quality of generated speech because the extra speech distortion after speech enhancement will propagate to the acoustic model as well as the vocoder~\cite{xie2021direct}. Motivated by denoising auto-encoder~\cite{vincent2008extracting}, denoising training strategy is also applied in several robust VC systems~\cite{mottini2021voicy, huang2021toward}. It has been reported that the denoising method leads to worse naturalness than the speech enhancement method, while the speech enhancement-based method has lower speaker similarity scores than the denoising approach~\cite{huang2021toward}. 

In the current two-stage waveform generation paradigm, Mel spectrogram is usually served as the bridge between the acoustic (conversion) model and the vocoder (waveform generator). But with the only noisy speech from target speakers, we cannot directly obtain a clean spectrum from the noisy waveform. The above methods either remove noise in advance or try to learn noise-invariant Mel spectrogram through a learning structure. Although Mel spectrogram is served as a bridge, there still exists a mismatch between the acoustic model and the vocoder as they are commonly separately learned and work on different distributions of speech representation.

In this paper, we leverage the advances of our recently proposed Glow-WaveGAN~\cite{cong2021glow} and propose a noise-independent speech representation learning approach for high-quality voice conversion for noisy target speakers. Specifically, instead of extracting Mel spectrogram for reconstructing the waveform, we propose to learn another kind of acoustic feature in the latent space where we ensure that the target distribution modeled by the conversion model is exactly from the modeled distribution of the waveform generator. With this premise, we manage to make the latent feature to be noise-invariant. To this end, we propose a noise-controllable WaveGAN that directly learns the noise-independent acoustic representation from waveform by its encoder and conducts noise-control in the hidden space through a feature-wise linear modulation (FiLM)~\cite{perez2018film} module in its decoder. In the noise-controllable WaveGAN, the encoder part acts as a robust feature extractor to extract the noise-independent acoustic representation, while the decoder part manages to reconstruct either noisy or clean speech through the conditional constraint. As for the conversion model, importantly, we utilize a flow-based model to learn the distribution of noise-independent but speaker-related latent features from phoneme posteriorgrams (PPGs)~\cite{sun2016phonetic, park2020cotatron, zhao2020voice}. In this way, the proposed method can produce high-quality converted speech for the target speakers with only noisy training data.


 \vspace{-6pt}
\section{Method}  
\begin{figure}[h]
    \begin{center}
     \includegraphics[scale=0.35]{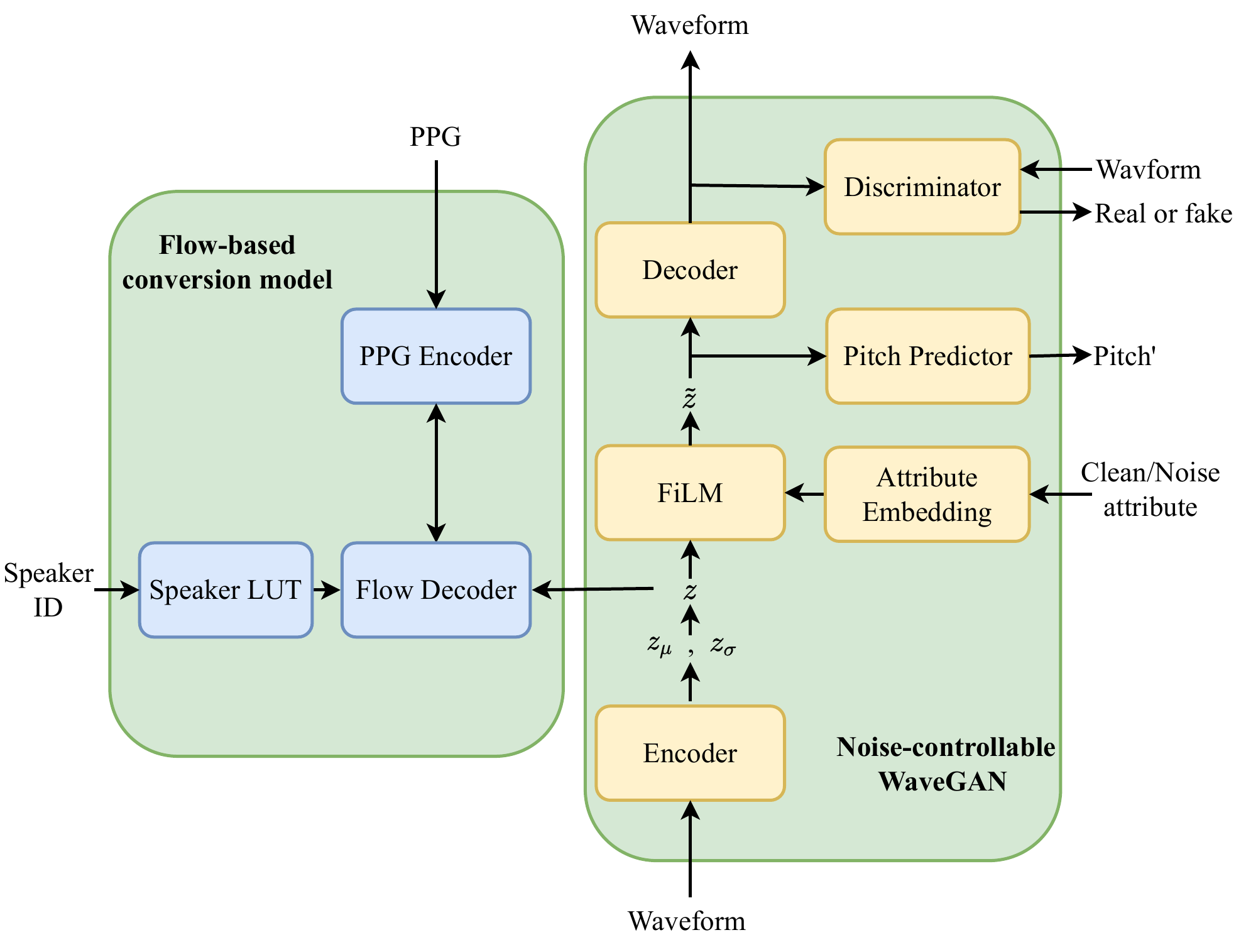}
    \end{center} \vspace*{-22pt} 
    \caption{The architecture of the proposed model.}
    \label{fig:model}	\vspace{-18pt}
\end{figure}

\vspace{-4pt}
The basic idea of the proposed voice conversion framework for noisy target speakers is to learn a \textit{noise-independent} acoustic representation of speech which bridges acoustic modeling and waveform generation. As shown in Figure~\ref{fig:model}, the proposed model comprises two major components: a noise-controllable WaveGAN module and a flow-based conversion module. Motivated by the Glow-WaveGAN model~\cite{cong2021glow}, which learns the hidden distribution of speech instead of conventional Mel spectrogram for high-quality speech synthesis, the noise-controllable WaveGAN (NC-WaveGAN) aims at learning noise-independent latent representation to remove the noise information from target speech for the down-stream voice conversion. The flow-based conversion model intends to model the distribution of the latent representation from phoneme posteriorgram (PPG). During inference, the conversion model maps the PPG of the source speaker's speech to the noise-independent latent representation of the target speaker, and the decoder part of NC-WaveGAN transforms the noise-independent representation to clean waveform. 

\vspace*{-15pt} 

\vspace{-5pt}
\subsection{Noise-controllable WaveGAN} \vspace{-3pt}
In our previous work~\cite{cong2021glow}, we proposed WaveGAN to learn the speech distribution in a compressed hidden space with an unsupervised manner for waveform reconstruction and down-stream speech synthesis tasks. The learning process is formulated as:  \vspace{-5pt}
\begin{equation}
   z = \text{Encoder}(w) \sim q(z|w)
\end{equation}  \vspace{-15pt}
\begin{equation}
   w = \text{Decoder}(z) \sim p(w|z)
   \label{eq:vae_dec}
\end{equation}
where $w$ is the waveform and $z$ is the latent representation. $q(z)$ is the hidden distribution of $w$ learned through unconditional VAE. In this formulation, the latent $z$ tends to include all aspects of input speech, like linguistic content, speaker and channel information. If the input speech $w$ is contaminated with noise, $z$ still contains the background noise in the noisy speech. 

To address this problem, we propose to learn a noise-independent representation $z$, which further benefits clean speech reconstruction for noisy target speaker in voice conversion. Specifically, to obtain noise-controllable WaveGAN (NC-WaveGAN), we use feature-wise linear modulation (FiLM)~\cite{perez2018film} to adaptively influence the encoder output by applying an affine transformation to the latent representation $z$. Here the FiLM takes a 2-dimensional attribute embedding $c$ as input. $c$ representing clean or noisy attribute of the waveform is defined as \vspace{-8pt}
\begin{equation}
c = 
\begin{cases}
(1, 0), & \text{if $w$ is clean}, \\
(0, 1), & \text{if $w$ is noisy}.
\end{cases}  \vspace{-3pt}
\end{equation}
where $c$ can be used a attribute vector to control the clean or noise condition.

Then the FiLM learns two linear functions $f$ and $h$ to modulate the input $z$ via a feature-wise affine transformation:  \vspace{-3pt}
\begin{equation}
  \widetilde z = f(c) \cdot z + h(c)   \vspace{-3pt}
\end{equation}
where $z$ is sampled from $q_(z)$ to compute the modulated $\widetilde z$ with the constraint of $c$. 

Note that in the original WaveGAN, it reconstructs waveform $w$ from the same input $w$. We argue that the paired clean and noisy speech share the same speech linguistic content and speaker identity, so the NC-WaveGAN reconstructs waveform $w'$ from the input waveform $w$ to make sure the clean/noise attribute is only from the FiLM module, where $w'$ has the opposite clean/noise attribute with $w$. So Equation~\eqref{eq:vae_dec} becomes \vspace{-7pt}
\begin{equation}
   w' = Decoder(\widetilde z) \sim p(w'|\widetilde z),
   \label{eq:vae_dec2}
\end{equation}
where $\widetilde z$ is modulated from the sample of $q(z|w)$. In this way, we treat the un-modulated $z$ as noise-independent speech representation. 

Since the Kullback-Leibler divergence is also evaluated between $q(z|w)$ and a prior $p(z)$, the training strategy of the NC-WaveGAN is similar to the original WaveGAN~\cite{cong2021glow}:  \vspace{-3pt}
\begin{equation}
    L_{vae} = L_{recon} + D_{KL}(q(z|w)||p(z))
    \label{eq:vae_loss}  \vspace{-3pt}
\end{equation} 
where $L_{recon}$ is the multi-resolution STFT loss~\cite{yamamoto2020parallel, binkowski2019high, yamamoto2019probability}, and $ D_{KL}(q(z|w)||p(z))$ is the Kullback-Leibler divergence. Besides, the additional pitch predictor and adversarial training are also applied in the NC-WaveGAN to improve the quality of reconstructed speech. Finally, the loss used to optimize NC-WaveGAN is  \vspace{-6pt}
\begin{equation}
    L = \lambda _{1} L_{vae} + \lambda _{2} L_{pitch} + \lambda _{3} L_{adv}.
   \vspace{-3pt}
\end{equation}
where $L_{pitch}$ and $ L_{adv}$ are the pitch reconstruction loss and the adversarial loss, respectively. Here $\lambda _{1}, \lambda _{2}$ and $\lambda _{3}$ are set empirically as in the original WaveGAN~\cite{cong2021glow}. 

\vspace{-5pt}
\subsection{Flow-based conversion module}\vspace{-3pt}
With the above NC-WaveGAN, we obtain a robust feature extractor to extract noise-independent latent $z$ and a vocoder that can control the noise/clean attribute of generated speech. When conducting voice conversion for noisy target speakers, we can directly model the noise-independent representation $z$ rather than the previously adopted noisy spectrogram. 

In this work, the conversion module models the target noise-independent distribution $q(z)$ from phoneme posteriorgram (PPG). Note that several generative models can model $q(z)$ from PPG, we choose the state-of-the-art flow-based generative model to maximize the likelihood of $q(z)$ by using invertible transformations~\cite{kingma2018glow,kim2020glow}. In practice, we simply follow the architecture of Glow-TTS~\cite{kim2020glow} to estimate $q(z)$, where the monotonic alignment search module is removed since the PPG and the latent $z$ have the same frame rate here. Besides, we also extend it to a multi-speaker scenario using a speaker look-up table (LUT), where speaker embedding is injected into the flow decoder.

The modeling and conversion process of the proposed model for a noisy target speaker can be summarized as follows.
\begin{enumerate}
	\item Given paired clean and noisy speech corpus from non-target speakers, we first train an NC-WaveGAN to obtain a feature encoder for noise-independent speech representation $z$ extracting, and also a decoder that reconstructs noisy or clean waveform from $z$.
	\item Given the noisy speech of the target speaker, we then build the flow-based conversion model with the corresponding PPGs extracted from an individual speech recognition model and $z$ from the trained encoder of NC-WaveGAN.
	\item Given the source speech during conversion, we extract the PPG and then generate $z$ from the flow model with the target speaker ID, and set the control vector in FiLM to ``clean'' to obtain the clean speech of the target speaker through the decoder of NC-WaveGAN.
\end{enumerate}
  
   \vspace{-8pt}
\begin{table}[h] 
  \caption{Settings of the training set used in three VC models and two neural vocoders.} \vspace{-5pt}
  \label{tab:data_settings}
  \centering
  \begin{tabular}{l ll l}
    \toprule 
     \multirow{2}{*}{\textbf{Models}}   & \multicolumn{2}{c}{\textbf{VCTK} }            &        \multirow{2}{*}{\textbf{Real-noise} }  \\ \cline{2-3} 
                                        & \textbf{VCTK-clean} & \textbf{VCTK-noise}                   \\
            \midrule
             Topline                 & clean                   & clean                         &\texttimes                \\
             Baseline                &  clean                 & denoised                      & denoised                                          \\
             FlowVC                    & clean                  & noisy                    &  noisy                                \\  \cline{1-4} 
             HiFi-GAN                   & clean                  & clean                         & \texttimes                                    \\
             NC-WaveGAN                 & clean+noisy            & \texttimes                          & \texttimes                                          \\
    \bottomrule
  \end{tabular} \vspace{-15pt}
\end{table} 
 
 \begin{table*}[t] \vspace{-5pt}
  \caption{Results of WER, CER and MOS scores for speech naturalness and speaker similarity with 95\% confidence intervals.} \vspace{-6pt}
  \label{tab:all_res}
  \centering
 \resizebox{\textwidth}{!}{
  \begin{tabular}{l | ll | ll | ll | ll }
    \toprule
      \multirow{2}{*}{\textbf{Model}}       & \multicolumn{2}{c|}{\textbf{\textbf{WER} $\downarrow$} }  & \multicolumn{2}{c|}{\textbf{CER} $\downarrow$ }     & \multicolumn{2}{c|}{\textbf{Naturalness MOS} $\uparrow$}      & \multicolumn{2}{c}{\textbf{Speaker similarity MOS} $\uparrow$ }     \\ \cline{2-9} 
      \      & \textbf{VCTK-noise}       & \textbf{Real-noise}       & \textbf{VCTK-noise}       & \textbf{Real-noise}      & \textbf{VCTK-noise}       & \textbf{Real-noise}     & \textbf{VCTK-noise}       & \textbf{Real-noise}   \\   
    \midrule                                    
    Topline                 & \textbf{14.32}      &  -                    & \textbf{4.99}       & -                             & 3.35 $\pm$ 0.07        & -                                    & 3.49 $\pm$ 0.07       & -                              \\
    Baseline                & 21.02               & 19.55                 & 7.83                & 6.76                          & 3.22  $\pm$ 0.07       & 2.98 $\pm$ 0.08                      & 3.41 $\pm$ 0.07        & \textbf{3.12 $\pm$ 0.09}       \\
    FlowVC                  & 19.77               & \textbf{14.25}        & 6.03                & \textbf{4.76}                 & \textbf{3.80 $\pm$ 0.05}       & \textbf{3.69 $\pm$ 0.06}     & \textbf{3.62 $\pm$ 0.07}        & 2.97 $\pm$ 0.10          \\
    \bottomrule
  \end{tabular}
}
\vspace{-3pt}
\end{table*}
 	\begin{figure*}[t]  \vspace{-2pt}
		\subfigure[Topline]{
			\begin{minipage}{0.32\linewidth}
				\centerline{\includegraphics[scale=0.22]{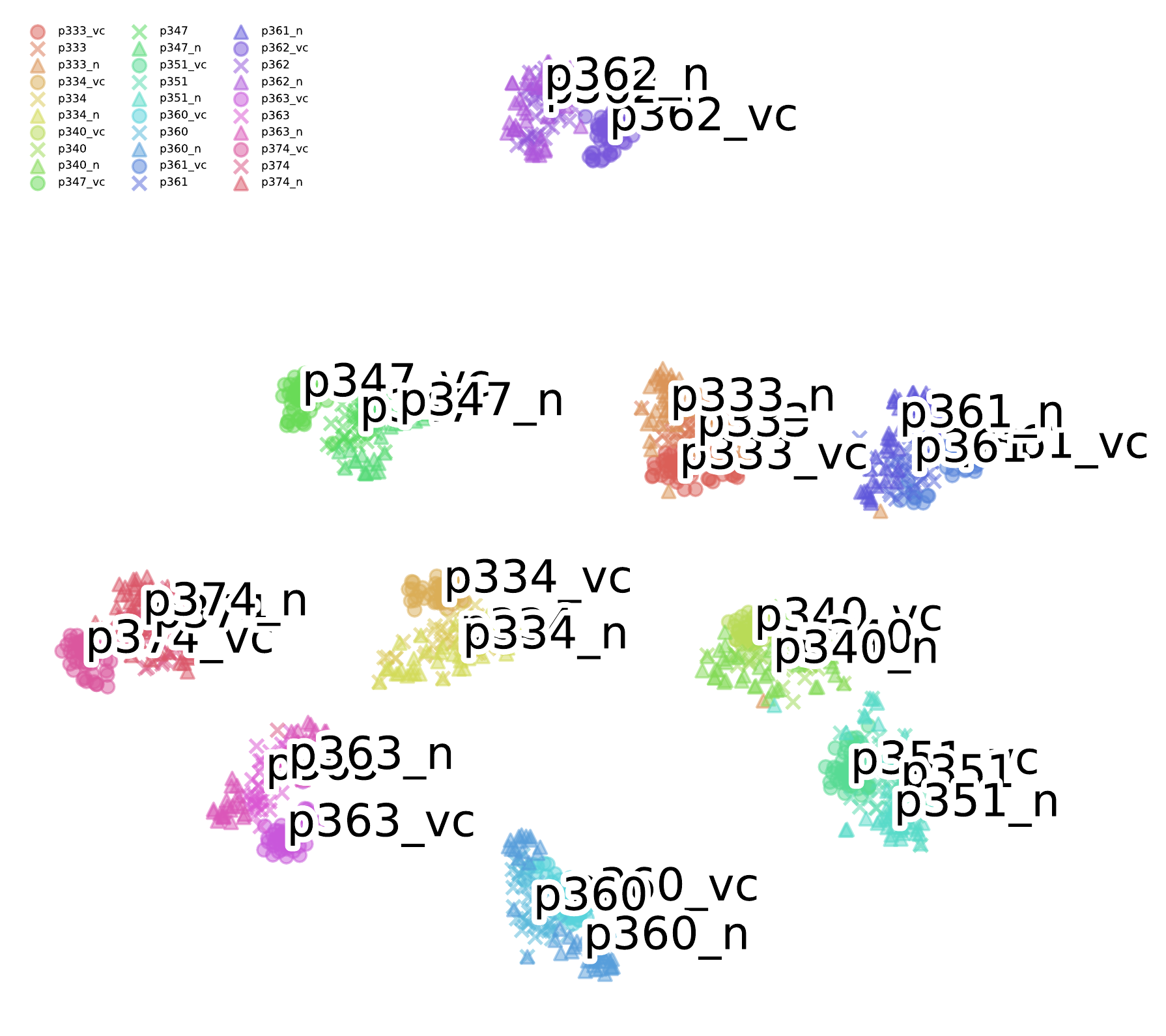}} \vspace{-2pt}
		\end{minipage}}
		\hfill \vspace{-2pt}
		\subfigure[Baseline]{ 
			\begin{minipage}{0.32\linewidth}
				\centerline{\includegraphics[scale=0.22]{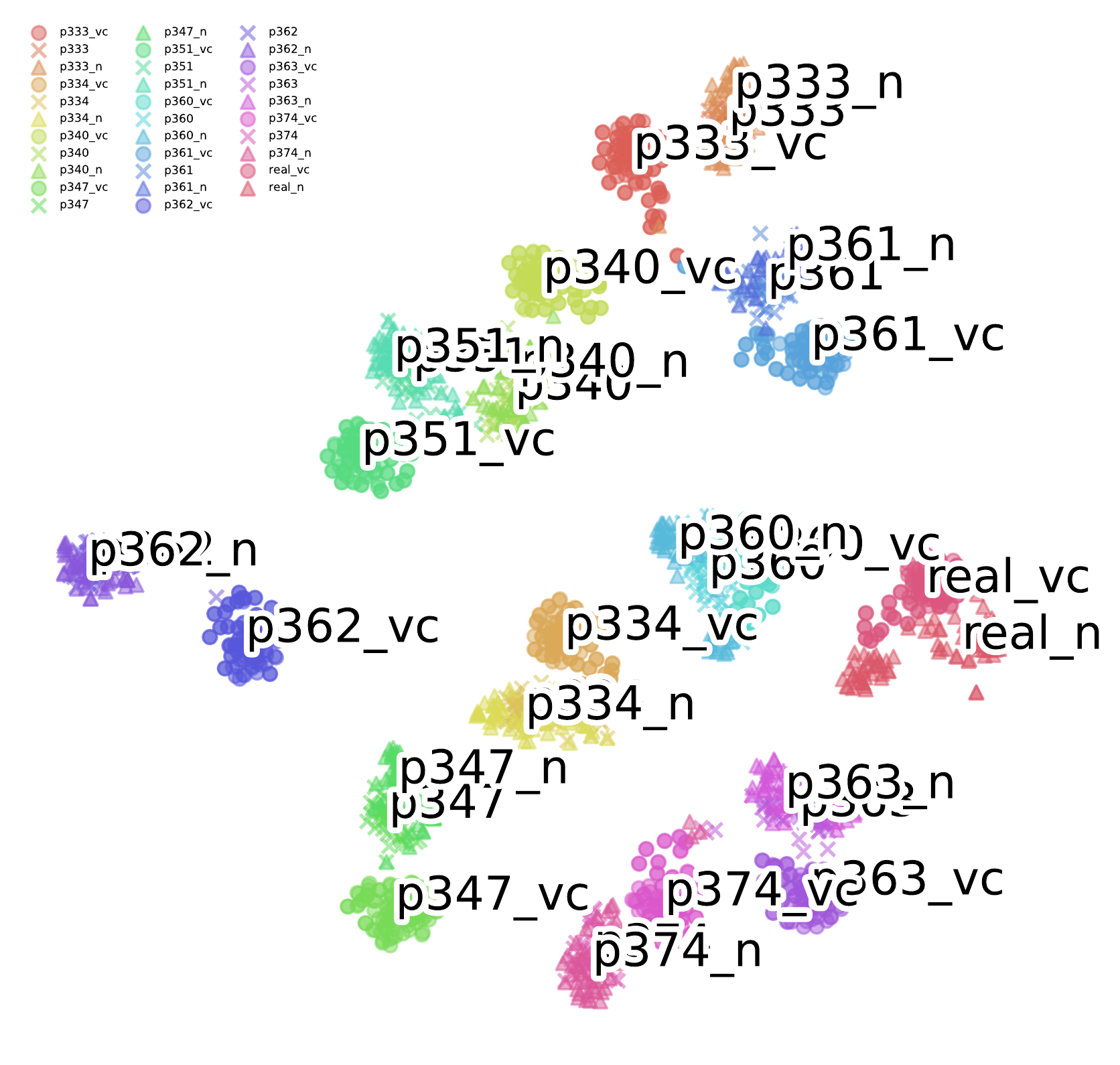}} \vspace{-2pt}
		\end{minipage}}
		\hfill \vspace{-2pt}
		\subfigure[FlowVC]{
			\begin{minipage}{0.32\linewidth}
				\centerline{\includegraphics[scale=0.22]{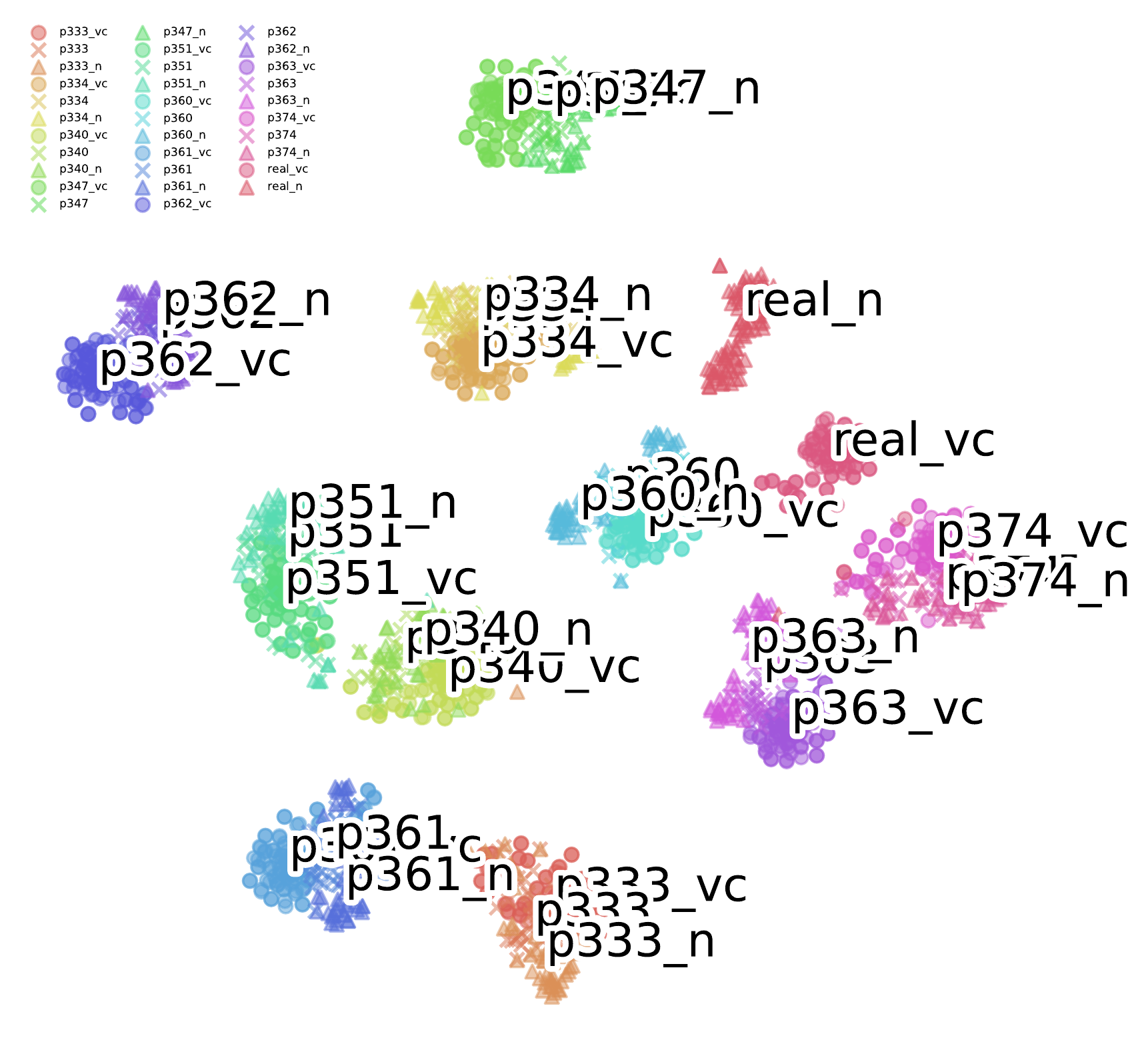}} \vspace{-2pt}
		\end{minipage}}
		\hfill  \vspace{-7pt}
		\caption{Visualization of speaker spaces learned by different models. The labels with the prefix of 'p' and 'real' denote the speakers of VCTK-noise and Real-noise, respectively. The labels with the suffix of '\_n' and  '\_vc' denote noisy utterances and voice conversion utterances, respectively. The labels without suffix denote clean utterances. } \vspace{-5pt}
		\label{fig:tsne} \vspace{-13pt}
	\end{figure*}

\vspace{-3pt}
\section{Experiments} \vspace{-2pt}
\subsection{Dataset and experimental setup} \vspace{-3pt}
To obtain paired clean and noisy speech for NC-WaveGAN training, we augment the VCTK corpus~\cite{Veaux2016SUPERSEDEDC} with noise from the CHiME-4 challenge~\cite{vincent2017analysis}. The VCTK corpus contains 44 hours of speech spoken by 109 English speakers, while the CHiME-4 corpus contains 8.5 hours of noise recorded in four different scenarios, including bus, cafe, pedestrian area, and street junction. We randomly select 55 speakers as clean speakers and refer to them as \textit{VCTK-clean}, while the rest 54 speakers' speech is mixed with random sampled noise at SNR ranging from 5 to 25 dB, which is referred to as \textit{VCTK-noisy}. We treat the speakers in the \textit{VCTK-noisy} set as the noisy target speakers in VC, while their clean references are NOT used during training in the proposed model. Apart from the simulated noisy target speakers, we also recruit another noisy speaker in-the-wild, named \textit{Real-noise}, to evaluate our approach in a real recording condition. The \textit{Real-noise} set contains 530 speech utterances, which contain noticeable noise and slight reverberation. We down-sample all audios into 16 kHz and use a 10ms frame shift for speech representation and PPG extraction. PPGs are extracted using an open-source speech recognition toolkit called WeNet~\cite{zhang2021wenet}, where we use the pre-trained WeNet model on the Librispeech corpus~\cite{panayotov2015librispeech} in this work.

To validate the performance of the proposed model, we train a topline model and a baseline model, both of which adopt a typical encoder-decoder-based auto-regressive voice conversion model mapping PPG to Mel spectrogram. And the HiFi-GAN vocoder~\cite{kong2020hifi} is adopted to reconstruct the waveform from the predicted Mel spectrogram. The auto-regressive voice conversion model contains a CBHG~\cite{lee2017fully} encoder and an auto-regressive decoder~\cite{shen2018natural}. As for the proposed model, the basic encoder and decoder of NC-WaveGAN follow the skeleton of WaveGAN~\cite{cong2021glow}, while the flow-based acoustic model is similar to the Glow-TTS~\cite{kim2020glow}.

Table~\ref{tab:data_settings} lists the settings of the training set used in three VC models, including topline, baseline and FlowVC, and two neural vocoders, including HiFi-GAN and NC-WaveGAN. The topline conversion model and the HiFi-GAN are trained with the original VCTK corpus (\textit{VCTK-clean} and the clean version of \textit{VCTK-noise}), which shows the upper limit of quality for the clean target speakers. Since the clean data of the \textit{Real-noisy} set is not available, we will not evaluate the topline for the real-world speaker. As for the baseline model, we directly conduct speech enhancement through the state-of-the-art pre-trained Uformer~\cite{fu2021uformer} model on all noisy target speakers to obtain denoised speech. For the proposed NC-WaveGAN, we only use the clean and the noisy version of \textit{VCTK-clean}, where the target noisy speakers in \textit{VCTK-noisy} and \textit{Real-noise} are excluded since the WaveGAN model is robust for unseen speakers. For the flow-based conversion model, referred to as FlowVC, the clean set \textit{VCTK-clean} and noisy target sets \textit{VCTK-noisy} and \textit{Real-noisy} are used for training.

\vspace{-8pt}
\subsection{Experimental results and analysis}  \vspace{-3pt}
We measure the speech quality and speaker similarity objectively and subjectively to evaluate all models. In each evaluation, we randomly select 50 source speech utterances and then convert them into clean speech for 11 noisy target speakers, including the target in \textit{Real-noisy} and 10 target speakers from \textit{VCTK-noisy}. For objective evaluation, we calculate word error rate (WER) and character error rate (CER) to measure the intelligibility of the converted speech. For subjective evaluation, we conduct mean opinion scores (MOS) to measure speech quality and speaker similarity. A group of 15 listeners who are proficient in English participates in the listening tests. The audio samples can be found on our demo page~\footnote{\url{https://lmxue.github.io/FlowVC/}}. Additionally, speaker space and speech representation are visualized to show speaker similarity and noise robustness of the proposed NC-WaveGAN. 

\textbf{WER and CER for intelligibility evaluation.} The WER and CER measured by the WeNet speech recognizer is shown in Table~\ref{tab:all_res}. Since the topline is adopted to show the upper limit of quality for clean speakers, we only evaluate its performance on the target speakers in the clean version of \textit{VCTK-noise}, where it achieves the lowest WER and CER values. As for the baseline model, it shows the highest WER and CER, which may be attributed to the distortion of the speech caused by speech enhancement. Meanwhile, the FlowVC gets significantly better results than the baseline on both simulated noisy targets (\textit{VCTK-noise}) and the noisy target speaker in-the-wild (\textit{Real-noise}), indicating that the proposed model can generate converted speech with better intelligibility for noisy target speakers.
%

\textbf{MOS for speech naturalness evaluation.} The naturalness of the converted noisy target speakers in terms of MOS is shown in Table~\ref{tab:all_res}. For the noisy target speakers in \textit{VCTK-noise}, we found the proposed method achieves significantly a higher score than the baseline model. We conjecture that it is because the denoised speech in the baseline may contain apparent distortion, which affects the perceived quality of the converted speech. Besides, since the flow-based conversion model directly models the distribution of the latent representation learned from NC-WaveGAN, which tends to eliminate the mismatch between the conversion model and the vocoding part, the proposed method outperforms the topline model that is even built on the clean speech of the target speakers. As for the noisy targets in the \textit{Real-noise} set, the proposed model still outperforms the baseline model significantly. Besides, we find the performance of the \textit{Real-noise} set from both the baseline and proposed model declines compared to that of \textit{VCTK-noise}, which may be caused by the reverberation of the target speaker in \textit{Real-noise} set. 

\textbf{Speaker similarity and visualization.} We also conduct subjective speaker similarity evaluation for the noisy target speakers, as shown in Table~\ref{tab:all_res}. For the \textit{VCTK-noise} set, although the target noisy speakers are excluded in the training of NC-WaveGAN, the proposed method still achieves the highest similarity score compared to other models, which proves the robustness of NC-WaveGAN for unseen speakers. But for the noisy target in \textit{Real-noise}, we find the similarity of the proposed model is worse than the baseline model.  Since the real noisy data of \textit{Real-noise} set also has reverberation, the pre-trained speech enhancement model (Uformer) also conducts dereverberation along with denoising, which removes noise and reverberation at the same time in the pre-processing stage. We find that dereverberation benefits to improving the speaker similarity of converted speech. While in NC-WaveGAN, the training data only contains additive noise, so the reverberation in the latent representation affects the performance of speaker similarity in speech reconstruction. 

To better analyze speaker similarity, we extract speaker embeddings of both original and converted speech through a publicly available pre-trained speaker verification (SV) system~\footnote{\url{https://github.com/resemble-ai/Resemblyzer}}, and then visualize them with t-SNE~\cite{van2008visualizing}, as shown in Figure~\ref{fig:tsne}. The visualization includes noisy speech, corresponding clean speech, and the converted speech for 11 testing target speakers. Each point in the visualization corresponds to a speech utterance, where different colors indicate different speakers and attributes (i.e., clean, noisy, or voice conversion result) labels. In Figure~\ref{fig:tsne} (a) for the topline model, different speakers are clearly separated, and the clean, noisy and converted utterances are overlapped, which demonstrates that the speaker embeddings have similar distribution among clean recordings, noisy recordings and converted speech. As for the baseline and proposed model, we can see clear speaker clusters of the FlowVC on the VCTK testing noisy speakers, while there exist clear boundaries between converted speech and the original recordings in the baseline model. Caused by the reverberation of the target speaker in \textit{Real-noise} set, there is an obvious distance between the converted speech labeled as ``real\_vc'' and the recordings labeled as ``real\_n'' in Figure~\ref{fig:tsne} (c), which inspires us to further investigate the reverberation condition in our model to meet more real application scenarios.

	\begin{figure}[h]
		\subfigure[$z$, before FiLM]{
			\begin{minipage}{0.43\linewidth}
				\centerline{\includegraphics[scale=0.21]{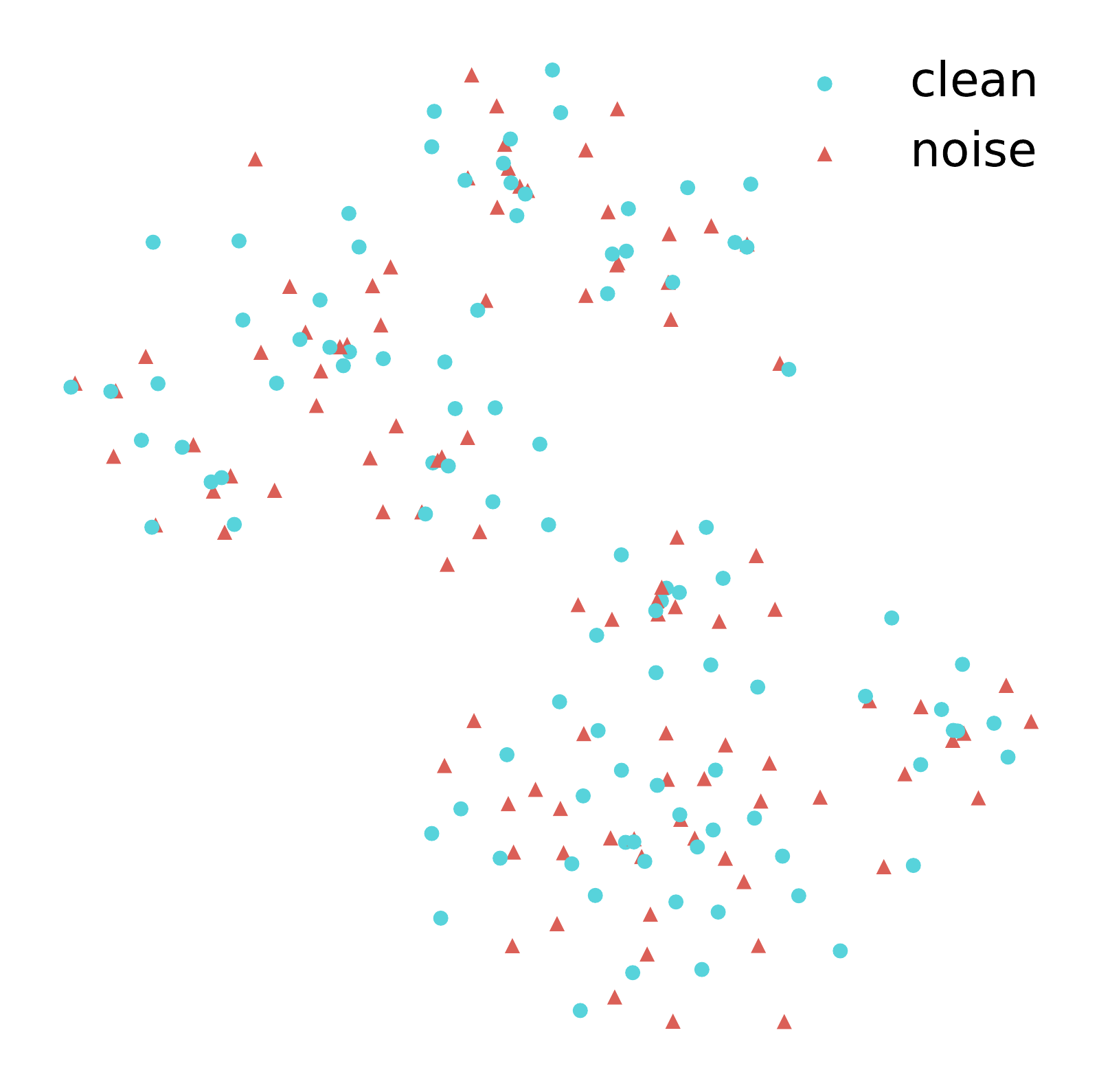}} 
		\end{minipage}}
		\hfill
		\subfigure[$\widetilde z$, after FiLM]{
			\begin{minipage}{0.43\linewidth}
				\centerline{\includegraphics[scale=0.21]{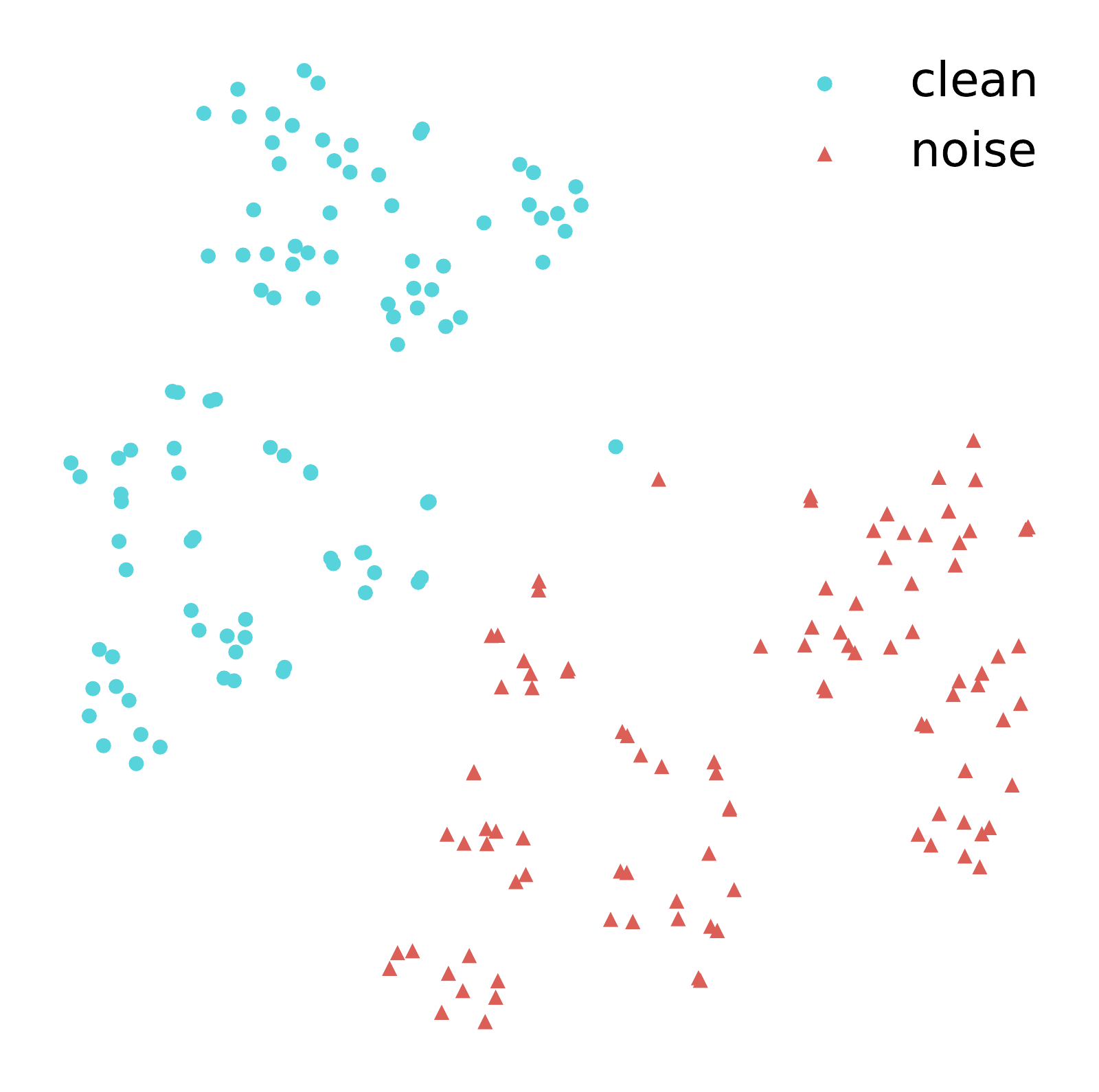}} 
		\end{minipage}}
		\hfill  \vspace{-10pt}
		\caption{Visualization of speech representations. }  
		\label{fig:z_w_o_film} \vspace{-22pt}
	\end{figure}
	
\textbf{Analysis of noise-independent representation} To confirm the learned acoustic representation is noise-independent, we randomly select 5 clean and 5 noise utterances from each of 10  \textit{VCTK-noise} speakers, totaling 100 utterances, which are then fed into NC-WaveGAN to generate 100 clean and 100 noisy speech utterances by two different FiLM control attributes, i.e., clean and noise. During generation, we extract the latent acoustic representations $z$ and their modulation version $\widetilde z$ after FiLM, and visualize them via t-SNE~\cite{van2008visualizing}, as shown in Figure~\ref{fig:z_w_o_film}. It is observed that before FiLM, $z$ from the clean and noisy speech are mixed together, while there exists a clear boundary among the FiLM modulated $\widetilde z$. It indicates that the proposed NC-WaveGAN can effectively learn noise-independent representation, and the FiLM can control the clean/noise attribute when reconstructing the waveform.

\vspace{-8pt}
\subsection{Summary}   \vspace{-4pt}
This paper proposed a noise-independent speech representation learning approach for high-quality voice conversion for noisy target speakers. Specifically, a noise-controllable WaveGAN is introduced to learn a noise-independent speech latent representation by its encoder, and reconstruct clean speech through a conditional constraint by its decoder. A flow-based conversion model is subsequently used to model the distribution of the speech representation learned by the noise-controllable WaveGAN. By using the acoustic feature in the latent space instead of Mel spectrogram, the proposed model directly models the distribution of the noise-independent speech representation, which eliminates the mismatch between the conversion stage and the waveform generation stage. Objective and subjective experiments demonstrate that the proposed model performs high speech quality and speaker similarity in voice conversion for noisy target speakers.
\bibliographystyle{IEEEtran}
\bibliography{main}

\end{document}